\documentclass[conference]{IEEEtran}
\ifCLASSINFOpdf
\else
\fi

\usepackage{tabstackengine}
\usepackage{cite}
\usepackage{multicol,lipsum}
\usepackage{amsmath,amssymb,amsfonts}
\usepackage{graphicx}
\usepackage{textcomp}
\usepackage{xcolor}
\usepackage{booktabs}
 \usepackage{hyperref}
\usepackage{enumerate}
\usepackage{multirow}
\usepackage{standalone}
\usepackage{tikz}
\usetikzlibrary{positioning}
\usetikzlibrary{decorations.markings}
\usepackage{float}
 \usepackage{makecell}
\usepackage{subcaption}
\usepackage{arydshln}

\makeatletter
\newcommand{\Biggg}{\bBigg@{4}}
\newcommand{\Bigggg}{\bBigg@{5}}
\makeatother

\begin{document}
%
\title{  Single Sparse Graph Enhanced Expectation Propagation  Algorithm Design  for Uplink MIMO-SCMA }
%
%
%

\author{Qu  Luo\textsuperscript{*},   Jing Zhu\textsuperscript{*},  Gaojie Chen\textsuperscript{*}, Pei Xiao\textsuperscript{*}  and Rahim Tafazolli\textsuperscript{*}  \\

\textsuperscript{*}  University of Surrey, Guildford,  GU2 7XH, Surrey, U. K. \\
Email: \{q.u.luo, j.zhu,  gaojie.chen, p.xiao, r.tafazolli\}@surrey.ac.uk.   
}
\maketitle

\begin{abstract}
 Sparse code multiple access (SCMA) and multiple input multiple output (MIMO) are considered as two efficient techniques to provide both massive connectivity and high spectrum efficiency for future machine-type wireless networks.  This paper proposes a single sparse graph (SSG) enhanced expectation propagation algorithm  (EPA) receiver, referred to as SSG-EPA, for uplink MIMO-SCMA systems. Firstly,  we reformulate the sparse codebook mapping process using a linear encoding model,  which transforms   the variable nodes (VNs) of SCMA
  from symbol-level   to bit-level VNs.  Such transformation facilitates     the integration   of the VNs of   SCMA    and low-density parity-check (LDPC), thereby  emerging the SCMA     and LDPC graphs into a  SSG.   Subsequently, to further reduce the detection complexity,  the message propagation between SCMA VNs and function nodes (FNs) are designed based on  EPA principles.   Different from the existing iterative detection and decoding (IDD) structure, the proposed EPA-SSG allows a simultaneously detection and decoding at each iteration, and eliminates the use of interleavers, de-interleavers, symbol-to-bit, and bit-to-symbol LLR transformations.  Simulation results show that the proposed SSG-EPA     achieves    better error rate performance compared to the state-of-the-art schemes.

\end{abstract}


%
\IEEEpeerreviewmaketitle

\section{Introduction}
 
\IEEEPARstart{N}{on-orthogonal} multiple access (NOMA) has been envisioned as a promising multiple access technique to support diverse traffic  with  stringent  requirements, such as  high  spectral efficiency, massive connectivity  and high-level quality of service (QoS),   for the   beyond fifth generation (B5G) and sixth-generation (6G) networks \cite{YuSparseStandards}.  
By  enabling overloaded transmission of multiple users,  existing NOMA techniques can be primarily categorized into power-domain NOMA  \cite{LuoNOMA,THZnoma}  and code-domain NOMA (CD-NOMA) \cite{CDNOMA}.  
Sparse code multiple access (SCMA) is a representative CD-NOMA scheme which can achieve maximum-likelihood decoding performance with low complexity \cite{LuoError}.    Furthermore, multiple-input multiple-output (MIMO) has been acknowledged as a   enabling technique for enhancing spectral efficiency by leveraging spatial multiplexing or diversity. Consequently, the integration of SCMA with MIMO, known as MIMO-SCMA, has gained increasing attention due to its enhanced spectral efficiency and error rate performance \cite{MIMOSCMADlink1,MengEPASCMA,MIMOSCMAEPQ2}.

By utilizing the sparsity structure in SCMA, message passing  algorithm (MPA) has been employed to achieve near-optimal error rate performance with lower complexity compared to maximum likelihood detection \cite{GCSCMA}.  
However, the complexity of  MPA  increases exponentially with the codebook size $M$ and the colliding users at each resource node. Therefore, many low-complexity detection approaches has also been proposed for SCMA \cite{MengEPASCMA,MIMOSCMAEPQ2}. A notable example is the expectation propagation  algorithm (EPA) for low-complexity detection in  MIMO-SCMA.  The use of  EPA for uplink  MIMO-SCMA detection was initially introduced in  \cite{MengEPASCMA}.  Later on, 
in \cite{MIMOSCMAEPQ2}, an EPA   structure  was developed for MIMO-SCMA with multi-antenna users based on the extended factor graph, where the computational   complexity was further reduced   by introducing QR decomposition and resource edge cluster-based decentralized factor node processing.   In addition, an approximate EPA  was proposed in \cite{EPA3}, where the message passing at the variable   and function nodes,  and the log likelihood ratio calculation were simplified to reduce the detection complexity.  The authors   \cite{DecentralizedMIMOSCMA} proposed a decentralized   processing scheme for low-complexity  detection in uplink massive MIMO-SCMA systems by utilizing the EPA framework.  The main idea is to partition the base station antennas into multiple independent antenna clusters.

 \begin{figure*}[htbp]
     \centering
     \includegraphics[width=0.79\linewidth]{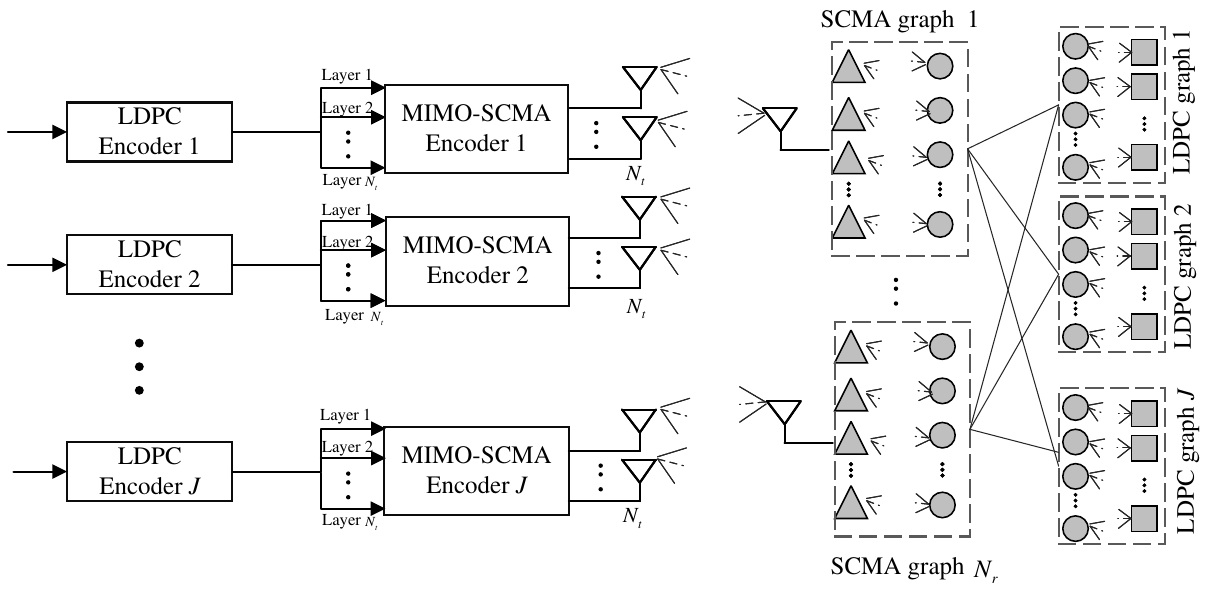}
     \caption{System model of uplink MIMO-SCMA.}
     \label{MIMO_scmaSysl}
     \vspace{-1em}
 \end{figure*}

Despite a  rich body of works on low-complexity detection design, there is a limited focus on the joint optimization of detection and decoding design for MIMO-SCMA systems.  The turbo like iterative detection and decoding  (IDD), referred to as turbo-IDD, has been widely investigated   for coded SCMA systems \cite{BICMSCMA,ICCSCMA}.  Notable examples include \cite{BICMSCMA} and \cite{ICCSCMA}.   In the turbo-IDD scheme, the SCMA detector and channel decoder are iteratively executed with a certain number of inner iterations independently for each outer iteration. However, the performance of the  turbo-IDD is generally limited as the inefficient message passing between SCMA detector and channel decoder. Benefiting from the sparse structures of the SCMA factor graph and Tanner graph of LDPC, the joint detection and decoding (JDD) has been proposed for  enhancing error rate performance by fully leveraging system diversity and coding gains \cite{GCJDDSCMA,LinyunSCMA}. The JDD structure treats SCMA detector and turbo decoder as a joint
factor graph, thereby improving the error rate performance and  facilitating    convergence. For example, the authors in \cite{LinyunSCMA} proposed an EPA-aided JDD structure for uplink  MIMO-SCMA, which showed that the proposed JDD-EPA receiver achieves significantly lower
complexity and faster convergence rate than the  turbo-IDD receiver.  It is noted that the messages between the SCMA detector and channel decoder in the JDD-EPA receiver still propagate sequentially. 

In this paper, we propose a single sparse graph (SSG) enhanced EPA receiver, referred to as SSG-EPA,  for uplink MIMO-SCMA systems.  Unlike the JDD receiver,  the detection and decoding in the proposed SSG-EPA receiver are performed simultaneously and in a joint manner at each iteration, eliminating the use of interleavers, de-interleavers, symbol-to-bit, and bit-to-symbol LLR transformations.  By reformulating the codebook with a linear encoding model, the variable nodes (VNs) of SCMA can be transformed from symbol-level VNs to bit-level VNs. Consequently, the VNs of low-density parity-check (LDPC) codes and SCMA can be merged, leading to the proposed SSG.   In addition, the EPA has been proposed for low complexity detection and the message propagation  in the proposed EPS-SSG has been discussed in detail. 
Simulation results show that the proposed SSG-EPA is able to achieve  low-detection complexity, fast convergence rate, and  good bit   error rate performance
\footnote{\textit{Notations:} Bold uppercase, bold lowercase and lowercase letters denote matrices, vectors and scalars, respectively.    $\mathbb{C}^{k\times n}$ and $\mathbb{B}^{k\times n}$ denote the $(k\times n)$-dimensional complex and binary matrix spaces, respectively.   $\text{diag}(\mathbf{x})$ gives a diagonal matrix with the diagonal vector of $\mathbf{x}$. $(\cdot)^\mathcal T$ and $(\cdot)^\mathcal H$ denote the transpose and the Hermitian transpose operation, respectively.  $\mathcal{CN}(0,1) $ denotes the complex distribution with zero mean and unit variance.  }.

\section{System Model}
 
We consider an uplink MIMO-SCMA system with $J$ users multiplexed over $K$ resource nodes (RNs), e.g., the OFDM subcarriers. The SCMA overloading is defined as $\lambda =J/K$,  which is larger than $100\%$.  The base station (BS) is equipped with $N_r$ receive antennas and each user is  equipped with $N_t$ transmit antennas, hereafter, we use $N_t \times N_r$ to denote the MIMO configuration. Fig. \ref{MIMO_scmaSysl}    shows system model of an uplink SCMA. 
During   transmission, the input message data are first encoded with a channel code. Then, the encoded bits are divided into $N_t$ layers for multiplexing. 
The   SCMA encoder of user $j$      maps   the encoded bits to an SCMA codeword $\mathbf {x}_{j}^{(n_{\mathrm {t}})}\in \mathbb {C}^{K\times 1}$ selected from the layer-specific complex codebook on the $n_t$th transmit antenna, which is denoted as $\mathcal {X}_{j}^{n_t}  \in \mathbb {C}^{K \times M}$.   In this paper, we assume the codebook for each antenna is the same, thus the superscript is omitted.  The detailed mapping process will be discussed later. In a standard  SCMA with single-input  single-output, each codeword has only $ N$ nonzero entries and the positions of the non-zeros elements remain the same within a codebook. Such a sparse structure can efficiently represented by an indicator graph matrix, denoted by ${\bf{F}}$,  which illustrates the sharing of the RNs among multiple VNs. An element of  ${\bf{F}}$ is defined as ${f_{k,j}}$ which takes the value of $1$ if and only if  VN  $v_j$ is connected to   RN  $r_k$ and 0 otherwise. In this paper, the following basic indicator matrix with $K=4, J=6 $ and $N=2$ is considered \cite{SSDSCMA}:
 \begin{equation} 
 \label{Factor_46}
 \small
 {{\mathbf{F}}_{4\times 6}}=\left[ \begin{matrix}
   0  & 1 & 1 & 0 & 1 & 0  \\
   1 & 0 &1& 0 & 0 & 1\\
   0 & 1& 0 & 1 & 0 & 1 \\
   1 & 0 & 0 &1 & 1 & 0  \\
\end{matrix} \right].
  \end{equation}

The received vector  at the $n_r$th antenna of the BS is given by
\begin{equation} 
\label{SigSISO}
 \mathbf {y}^{n_{\mathrm {r}}}= \sum \limits ^{J}_{j=1}\sum \limits ^{N_{\mathrm {t}}}_{n_{\mathrm {t}}=1}\mathrm {diag}\left ({\mathbf {h}_{j}^{(n_{\mathrm {r}},n_{\mathrm {t}})}}\right)\mathbf {x}^{n_{\mathrm {t}}}_{j}+\mathbf {z}^{n_{\mathrm {r}}},
 \end{equation}
 where $n_{\mathrm {r}}=1,2,\ldots,N_{\mathrm {r}}$, and $\mathbf {h}^{(n_{\mathrm {r}},n_{\mathrm {t}})}_{j}\in \mathbb {C}^{K\times 1}$ is the   channel vector from the $n_t$th transmit antenna to the  $n_r$th receive antenna of user $j$, $\mathbf z^{n_r} \in \mathbb C ^{K \times 1}$ denotes the  independent additive white Gaussian noise (AWGN)   vector  at the $n_r$th antenna of the BS,  $\mathbf y^{n_r}=[y_{1}^{n_r},y_{2}^{n_r},\cdots,y_{K}^{n_r}]^{\mathcal T}$ denotes  the   received signal vector on the BS. Let $\mathbf {y}\equiv \left [{ \mathbf {y}^{1,{\mathcal T } }, \mathbf {y}^{2,{\mathcal T  } },\cdots,\mathbf {y}^{N_{\mathrm {r}},{\mathcal T  }}}\right]^{\mathcal T  }\in \mathbb {C}^{N_{r}J\times 1}$ collect the received signal from $N_r$ antennas, then  the received signal can be re-written as 
\begin{equation} 
\mathbf {y}= \sum \limits_{j=1}^{J}\mathbf {H}_{j}\mathbf {x}_j+\mathbf {z},
\end{equation}
 where $\mathbf {x}_j\equiv \left [{ \mathbf {x}_j^{1,{\mathcal T } }, \mathbf {x}_j^{2,{\mathcal T  } },\cdots,\mathbf {x}_j^{N_{\mathrm {t}},{\mathcal T  }}}\right]^{\mathcal T  }\in \mathbb {C}^{N_{t}J\times 1}$ denotes the transmitted signal vector, $\mathbf {z}\equiv \left [{ \mathbf {z}^{1,{\mathcal T } }, \mathbf {z}^{2,{\mathcal T  } },\cdots,\mathbf {z}^{N_{\mathrm {t}},{\mathcal T  }}}\right]^{\mathcal T  }\in \mathbb {C}^{N_{t}J\times 1}$ denotes the  complex AWGN noise vector, and 
 $\mathbf H_j \in \mathbb C^{N_rK\times N_tK}$ denotes the MIMO channel matrix from the $j$th UE to the  BS, which takes the following expression
 \begin{equation}
 \small
 \label{Hl}
\begin{aligned} \mathbf {H}_{j}\equiv \left [{ \begin{matrix} \mathrm {diag}(\mathbf {h}^{(1,1)}_{j}) &\quad \!\! \mathrm {diag}(\mathbf {h}^{(1,2)}_{j}) &\quad \!\! \cdots &\quad \!\! \mathrm {diag}(\mathbf {h}^{(1,N_{\mathrm {t}})}_{j}) \\ \mathrm {diag}(\mathbf {h}^{(2,1)}_{j}) &\quad \!\! \mathrm {diag}(\mathbf {h}^{(2,2)}_{j}) &\quad \!\! \cdots &\quad \!\! \mathrm {diag}(\mathbf {h}^{(2,N_{\mathrm {t}})}_{j}) \\ \vdots &\quad \!\! \vdots &\quad \!\! \ddots &\quad \!\! \vdots \\ \mathrm {diag}(\mathbf {h}^{(N_{\mathrm {r}},1)}_{j}) &\quad \!\! \mathrm {diag}(\mathbf {h}^{(N_{\mathrm {r}},2)}_{j}) &\quad \!\! \cdots &\quad \!\! \mathrm {diag}(\mathbf {h}^{(N_{\mathrm {r}},N_{\mathrm {t}})}_{j}) \\ \end{matrix} }\right]. \\ 
\end{aligned}
 \end{equation}

\section{The Proposed Single Graph Design }

\subsection{Preliminary } 

This section introduces the preliminaries   in MIMO-SCMA detection and decoding.  


\textit{1) Turbo receiver with iterative detection and decoding (Turbo-IDD) :} In the Turbo-IDD structure,  iterative messages are exchanged between the multiuser detector (MUD) and channel decoder. Denote $N_{\text{iter}}^{\text{MUD}}$ and $N_{\text{iter}}^{\text{DEC}}$ by the number of inner iterations of the MUD and  channel decoder, respectively. Similarly, we further denote  $N_{\text{iter}}^{\text{out}}$ as the number of outer iterations.  

\textit{2) Joint detection and decoding:} In an LDPC coded SCMA system, both MUD and channel decoder can be represented with sparse factor graphs.   By   merging the SCMA factor and LDPC graph into a integrated one,   the belief message are able to exchange efficiently over    an aggregated factor graph associated to both the sparse codebooks and LDPC codes.

Although the JDD shows significant performance improvement over the Turbo-IDD  receiver, the LLR exchange between the SCMA factor graph and LDPC graph makes belief message propagation inefficient. Specifically, the SCMA  VNs  need to transform the LLR of each codeword to bit-level LLRs and pass them to the deinterleaver. Similarly, the output bit-level LLRs from the LDPC VNs need to be transformed back to symbol LLRs. These LLR transformations and interleavers/deinterleavers prevent belief messages from propagating in a single graph manner. In this section, we present the proposed SSG design for MIMO-SCMA systems.

\subsection{
 Codebook Reformulation }

For  conventional SCMA codebook,  the  encoder maps the input binary message to a codeword selected from a pre-defined codebook codebook, i.e.,
\begin{equation} \label{SCMAmapping}
\small
f_{j}:\mathbb {B}^{\log _{2}M\times 1}\mapsto   \boldsymbol {{\mathcal X}}_{j}, \quad {~\text {i.e., }}\mathbf {x}_{j}=f_{j}(\mathbf {b}_{j}),
\end{equation}
where $\mathbf {b}_{j}=[b_{j,1},b_{j,2},\ldots,b_{j,\log _{2} M}]^{T}\in \{1,-1\}^{\log _{2} M}$ stands for $j$th user's  instantaneous input binary message vector, $   \boldsymbol {{\mathcal X}}_{j} \in \mathbb C^{K \times M}$ denotes the codebook of $j$th user   and $f_j$ denotes the SCMA encoding process. It should be noted that the codebook is deigned with symmetric. In fact, many of existing SCMA codebooks has such property, which empowers the SCMA encoding process can be written as a linear encoding process. Namely,   
\begin{equation} \label{design}
\small
\mathbf {x}_{j} =  {{\mathbf{G}}_{j}}{{\mathbf{b}}_{j}},
 \end{equation}
 where ${{\mathbf{G}}_{j}}\in {{\mathbb C}^{K\times \text{log}_2M}}$ is the codebook generator matrix of the  $j$th user. By collecting all the  $\mathbf {b}_{j}$ according to their corresponding integer values in ascending order, we form a $\log_2(M) \times M$ binary matrix $\mathbf{B}$. For example, when $M= 4$ and $16$, we have 
\begin{equation}
\label{U1}
      \mathbf {B} =\left [{ \begin{matrix} -+-+\\ --++ \end{matrix} }\right], 
\end{equation}
and 
\begin{equation}
\label{U2}
    \mathbf {B} =\left [{ \begin{matrix} -+-+-+-+-+-+-+-+\\ --++--++--++--++\\ ----++++----++++\\ --------++++++++ \end{matrix} }\right] ,  
\end{equation}
respectively, where ``$+$'' and ``$-$'' denote $+1$ and $-1$, respectively.  It should be noted that the  $m$th column of  (\ref{U1}) and (\ref{U2}) are symmetric with their negative values of $(M-m)$th column. Namely,   the values at $(M-m)$th column are equal to the negative of the value of $m$th column.
Hence, for the codebook that also has such  symmetric property, the codebook generator ${{\mathbf{G}}_{j}}$ can be generated as \cite{E2E}
\begin{equation}
   \mathbf  G_{j }  = \boldsymbol {{\mathcal X}}_{j}    \mathbf {B}^{\mathcal T}(\mathbf {B}\mathbf {B}^{\mathcal H})^{-1}.
\end{equation}
Hence, the signal model of (\ref{SigSISO}) can be rewritten as 
\begin{equation} 
\begin{aligned}
  \mathbf {y}^{(n_{\mathrm {r}})} =& \sum \limits ^{J}_{j=1}\sum \limits ^{N_{\mathrm {t}}}_{n_{\mathrm {t}}=1} \mathrm {diag}\left ({\mathbf {h}_{j}^{(n_{\mathrm {r}},n_{\mathrm {t}})}}\right) \mathbf  G_{j } \mathbf {b}^{(n_{\mathrm {t}})}_{j}+\mathbf {z}^{(n_{\mathrm {r}})} \\
   = & \sum \limits ^{J}_{j=1}\sum \limits ^{N_{\mathrm {t}}}_{n_{\mathrm {t}}=1}   {\overline{  \mathbf {H}}_{j}^{(n_{\mathrm {r}},n_{\mathrm {t}})}}    \mathbf {b}^{(n_{\mathrm {t}})}_{j}+\mathbf {z}^{(n_{\mathrm {r}})},
\end{aligned}
 \end{equation}
where $\overline{  \mathbf {H}}_{j}^{(n_{\mathrm {r}},n_{\mathrm {t}})}  =  \mathrm {diag}\left ({\mathbf {h}_{j}^{(n_{\mathrm {r}},n_{\mathrm {t}})}}\right) \mathbf  G_{j}$ denotes the effective channel matrix.
 
\subsection{The Proposed SSG-EPA}

For the MIMO-SCMA system depicted in Fig. \ref{MIMO_scmaSysl}, the  joint distribution $ p(\mathbf b, \mathbf y)$ can be respectively factorized as
\begin{equation}
\begin{aligned}
   p(\mathbf y,\mathbf b)= &p(\mathbf y \vert \mathbf b)p(\mathbf b) \\=&  \prod _{n_t=1}^{N_t} \prod _{j=1}^{J} \prod _{m=1}^{\log_2 M}p( b_{j,m}^{n_t})  \prod _{n_r=1}^{N_{r}}\prod _{k=1}^{K} p(y^{n_r}_{k}\vert  b_{j,m}^{n_t}). 
\end{aligned}
\end{equation}

\begin{figure}
    \centering
    \includegraphics[width=0.7\linewidth]{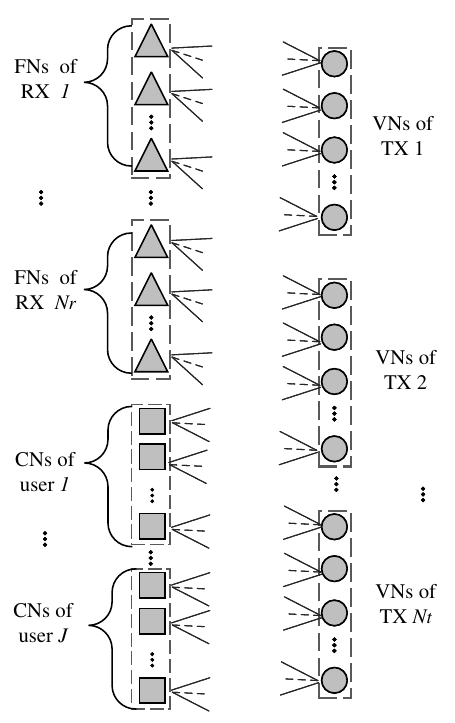}
    \caption{An example of the proposed SSG.}
    \label{SSG}
\end{figure}

The SCMA factor graph  can be designed to contain $JN_t\log_2M$ VNs   and $N_rK$ likelihood FNs. For easy of presentation, the SCMA VNs are denoted by $v_n^m, n= 1,2,\ldots, JN_t,  m=1, 2\ldots,\log_2M$ and the FN at the $k$th subcarrier of $n_r$th receiver antenna is denoted as   $f_{k}^{n_r}$. The corresponding factor graph representation of the proposed SSG is shown in Fig. \ref{SSG}.  
Note that the SCMA   and LDPC VNs  are all  propagate messages in bit-levels, hence, it is reasonable to place  SCMA FNs (triangles) and LDPC CNs (rectangles)  on one side, while drawing bit VNs (circles) on the other side.  The LLR calculation of  bit to symbol and symbol to bit are no longer required and the interleaver (deinterleaver)  can also be eliminated as it can be emerged within the SSG.   

In this paper, we consider the EPA as the MUD for low complexity detection. EPA is a Bayesian inference method designed to approximate the  complex  posterior belief distribution using a tractable distribution, typically a Gaussian distribution achieved through moment matching.  Define the projection of a distribution $p$ into a distribution set $\Phi$ as \cite{MIMOSCMAEPQ2}
\begin{equation} {\mathrm {Pro}}{{\mathrm {j}}_{\Phi } }\left ({p }\right) = \arg \min _{q \in \Phi } \text{KL}\left ({{p||q} }\right), \end{equation}
where $\text{KL}\left ({{p||q} }\right)$ denotes the Kullback Leibler (KL) divergence that measures the the similarity between the \textit{a posterior} belief distribution and the tractable distribution after projection.  Namely, we  have
\begin{equation}
\begin{aligned} 
\text{KL} (p|| q)= \sum _{\mathbf {b}}p\left ({\mathbf {b}|\mathbf {y}}\right)\log \frac {p\left ({\mathbf {b}|\mathbf {y}}\right)}{q\left ({\mathbf {b}|\mathbf {y}}\right)}.  
\end{aligned}
\end{equation} 
where $\mathbf {b}$ denotes the transmitted message bits of $J$ users.  Denote $   I_{v_{n}^{m}\rightarrow f_{k}^{n_r}}^{t}({b}_{n}^{m})$ and $ I_{f_{k}^{n_r} \rightarrow v_{n}^{m} }^{t}( {b}_{n}^{m}) $ as the belief message propagated from VN $v_{n}^{m}$  to  FN  $f_{k}^{n_r}$ and   FN  $f_{k}^{n_r}$  to  VN $v_{n}^{m}$at the $t$th iteration, respectively.   We  further denote  the set of SCMA   VNs  indices sharing the  $f_k^{n_r}$ as $F(k,n_r)$, whereas  the set of FN  indices sharing the $ v_n^m$ is denoted by $V(n)$.     According to the EP principles, the message update rule can be expressed as \cite{LinyunSCMA}:
\begin{equation}
\label{IV2F}
  I_{v_{n}^{m}\rightarrow f_{k}^{n_r}}^{t}({b}_{n}^{m})\propto\frac{\text{Proj}_{\Phi}(q_0^{~t}({b}_{n}^{m}) )} {I_{f_{k}^{n_r} \rightarrow v_{n}^{m} }^{t-1}( {b}_{n}^{m})}, 
\end{equation}
\begin{equation}
  I_{f_{k}^{n_r} \rightarrow v_{n}^{m} }^{t}( {b}_{n}^{m}) \propto \frac{\text{Proj}_{\Phi}(q_{k,{n_r}}^{t}({b}_{n}^{m}))}{ I_{v_{n}^{m}\rightarrow f_{k}^{n_r}}^{t}({b}_{n}^{m})},
\end{equation}
where  
\begin{equation}
    q_0^{~l}({b}_{n}^{m})          \propto p_0({b}_{n}^{m}) \prod _{n_r=1}^{N_{r}} \prod _{ k,n_r \in V(n)} I_{f_{k}^{n_r} \rightarrow v_{n}^{m} }^{t}( {b}_{n}^{m}),  
\end{equation}
\begin{equation}
\label{qnn}
   q_{k, {n_r}}^{t}({b}_{n}^{m})   \propto \sum_{  {b}_{\bar n}^{\bar m  } \neq {b}_{ n }^{ m }} p(y_{k}^{n_r} \vert {b}_{ \bar n}^{\bar m}) \prod_{ n,m  \in F(k,n_r)} I_{v_{n}^{m}\rightarrow f_{k}^{n_r}}^{t}({b}_{n}^{m}).
\end{equation}
 The above procedure can be understood as the    messages propagation over the  SCMA graph    depicted in Fig. \ref{SSG}. 
It treats the messages exchanged between  VNs  and    FNs  as continuous random variables and approximates the true distribution with a Gaussian distribution, characterized by its mean and variance. This approach enables signal detection to be transformed into the computation of mean and variance, avoiding the need to traverse all symbols. Specifically, we have
\begin{subequations}
\begin{align}
       \text{Proj}_{\Phi}(q_0^{~t}({b}_{n}^{m}) ) \propto & \mathcal {CN}({\mu _{n,m}^{t},\xi _{n,m}^{t}}),  \label{eq_a} \\ 
       I_{v_{n}^{m} \rightarrow  f_{k}^{n_r}}^{t}({b}_{n}^{m})\propto & \mathcal {CN}({\mu _{n,m\rightarrow  k,n_r}^{t},\xi _{m,t\rightarrow  k,n_r}^{t}}), \label{eq_b} \\ I_{  f_{k}^{n_r}\rightarrow v_n^m}^{t}({b}_{n}^{m}) \propto & \mathcal {CN}({\mu _{k,n_r\rightarrow n,m}^{t},\xi _{ k,n_r\rightarrow n,m}^{t}}). \label{eq_c}
\end{align}
\end{subequations}

 In the following, we present  the detailed message propagation in the proposed SSG.  Denote   $ c_{l}^j$ as the $l$th LDPC   check node (CN) of user $j$. For simplicity, the superscript $j$ is omitted. 

\subsubsection{Initialization}
Assuming there is no \textit{a priori} probability available, initial LLRs are set to zeros. Namely, 
\begin{equation}
     L_{  v_n^m  \rightarrow c_{i}} = 0,  \quad  L_{  c_{i}  \rightarrow v_n^m } = 0,   \quad \forall i, \forall n,  \forall m,
\end{equation}
where $   L_{  v_n^m  \rightarrow c_{i}} $ and $ L_{  c_{i}  \rightarrow v_n^m }$ denote  the LLRs propagating   from the  VN  $v_n^m $ to the $i$th LDPC CN  and the opposite direction, respectively. 

\subsubsection{Update of    VNs to CNs and FNs}
The updating of  VNs to CNs can be calculated as  
\begin{equation}
\label{LAB}
    L_{  v_n^m   \rightarrow c_{i}} =  \sum_{(k,n_r) \in V(n)}  L_{f_k^{n_r} \rightarrow v_n^m }   + \sum_{i^{\prime} \in \eta(n,m)  \setminus i } L_{c_{i^{\prime}} \rightarrow v_n^m},
\end{equation}
where $\eta(n,m)$ denotes the set of CNs connected to VN $v_n^m$ and $L_{f_k^{n_r} \rightarrow v_n^m }$ denotes the LLR propagated from the FN  $f_k^{n_r}$ to the VN $v_n^m$. 

To update the message from  VNs to FNs, each  VN  first gathers   the LLR  from both CNs and FNs, i.e.,
\begin{equation}
\label{LAF}
    L({ v_n^m   })=  \sum_{(k^{},n_r^{}) \in V(n) \setminus (k,n_r) 
 }  L_{f_{k^{}}^{n_r^{}} \rightarrow  v_n^m  }   + \sum_{i  \in \eta(n,m)      } L_{c_{i} \rightarrow  v_n^m }.
\end{equation}

  Upon obtaining  $  L({ v_n^m   })$,    the   mean and variance of the  VN $v_n^m$  can be calculated   as \cite{{LinyunSCMA}}
\begin{equation}
\begin{aligned}
       \mu _{n,m}^{t}& = 1- 2p( b_{n,m} =  -1 \vert \mathbf y)   \\
       &= \frac { 1- \exp \left (   L({ v_n^m   }) \right)}{{1 + \exp \left (    L({ v_n^m   }) \right)}}, \\
       \xi _{n,m}^{t}& =  \sum _{  u_{n,m} \in \{-1,1\} } p( b_{n,m}=  u_{n,m} \mid \mathbf y){\left |{u_{n,m} \! - \!\mu _{n,m}^{t}}\right |}^{2}.
\end{aligned}
\end{equation}
To proceed, denoted    $\mu _{n,m\rightarrow k,n_r}^{t}$ and $\xi _{n,m\rightarrow k,n_r}^{t}$ by the  mean  and variance  passing from VN $v_n^m$ to FN $f_{k^{}}^{n_r^{}}$ at $t $th iteration, respectively. 
Based on the EP principles, one has the    $\mu _{n,m\rightarrow k,n_r}^{t}$ and $\xi _{n,m\rightarrow k,n_r}^{t}$ updated as follows:
\begin{equation}
\small
\label{m_v_ejsg}
\begin{aligned}
        \xi _{n,m\rightarrow k,n_r}^{t}=&\left ({\frac {1}{\xi _{n,m}^{t} }-\frac {1}{\xi _{k,n_r\rightarrow n,m}^{t-1}}}\right)^{-1}, \\
        \mu _{n,m\rightarrow k,n_r}^{t}=& \xi _{n,m\rightarrow k,n_r}^{t} \left({\frac {\mu _{n,m}^{t} } { \xi _{n,m}^{t}}  -  \frac { \mu _{k,n_r\rightarrow n,m}^{t-1}} {\xi _{k,n_r\rightarrow n,m}^{t-1} 
 }}\right),
\end{aligned}
\end{equation}
  where $\mu _{k,n_r\rightarrow n,m}^{t}$ and $\xi _{k,n_r\rightarrow n,m}^{t}$ denote the  mean  and variance  passing from FN $f_{k^{}}^{n_r^{}}$ to  VN $v_n^m$  at $t $th iteration, respectively.
  
\subsubsection{Update of CNs and FNs to VNs}   
Based on   (\ref{qnn})and ({\ref{eq_a}})-({\ref{eq_c}}) , the message in the opposite direction becomes
\begin{equation}
\small
\label{Mw_va}
     \mu _{k,n_r\rightarrow n,m}^{t} =\frac {y_{k}^{n_r}-Z _{k,n_r\rightarrow n,m}^{t} }{h_{k,n_r}^{n,m}}, \; \xi _{k,n_r\rightarrow n,m}^{t}=\frac {N_0+ B_{k,n_r\rightarrow n,m}^{t}}{\left |h_{k,n_r}^{n,m}\right |^{2}},
\end{equation}
where 
\begin{equation}
\small
\begin{aligned}
    Z _{k,n_r\rightarrow n,m}^{t} =& \sum _{ \substack {i,j\in F(k,n_r), \\ i\neq n, j \neq m}} h_{k,n_r}^{i,j} \mu _{i,j\rightarrow k,n_r}^{t-1},  \\
    B_{k,n_r\rightarrow n,m}^{t}=& \sum _{ \substack {i,j\in F(k,n_r), \\ i\neq n, j \neq m}} \vert h_{k,n_r}^{i,j} \vert ^2\xi _{i,j\rightarrow k,n_r}^{t-1}.
\end{aligned}
\end{equation}
Finally, the posterior probability of transmitted symbol in LLR domain is updated by
\begin{equation}
  \small
  \label{F_V_EJSG}
      L_{f_n^r\rightarrow v_n^m}^{t}  =  \log\frac{\left |{1- \mu _{n,r\rightarrow m,t}^{t} } \right | ^{2}}  {\left |{1+\mu _{n,r\rightarrow m,t}^{t} } \right | ^{2}}.
  \end{equation}

 The LLR of the parity CN $c_i$ is updated as
\begin{equation}
\small
\label{CNupd1}
    L_{ c_{i}  \rightarrow v_n^m}=\gamma^{-1}\left(\sum_{(n^{\prime},m^{\prime})\in\phi_{i} \setminus (n,m)}\gamma\left(L_{ v_{n^{\prime}}^{m^{\prime}} \rightarrow c_i}\right)\right),
\end{equation}
where  $\phi_{i} \setminus (n,m)$ is the set of VNs (excluding $v_n^m$) that connect to the CN $c_{i}$, and $ \gamma(x) $ and $\gamma^{-1}(x)$ are respectively defined as 
\begin{equation}
\small
\label{CNupd2}
\begin{aligned}
     \gamma(x) &=\text{sign}(x)\times\left(-\log\tan{\vert x\vert\over 2}\right), \\
    \gamma^{-1}(x)& =(-1)^{\text{sign}(x)}\times\left(-\log\tan{\vert x\vert\over 2} \right),
\end{aligned}
\end{equation}
where $\text{sign}(x)$ denotes the the sign of $x$. 

\subsection{Computational Complexity}

 The computational complexity of EPA in the proposed SSG at each iteration  can be approximated by  $\mathcal O(N_{\text{SCMA}} N_r N d_f \log_2 M)$, where $N_{\text{SCMA}}$ denotes the number of SCMA symbols at each LDPC block.  Denoted  $d_{v}^{\text{ldpc}}$  where $d_{c}^{\text{ldpc}}$ by  the average column weight and the average row weight of the parity check
matrix, respectively. Consequently, the complexity of LDPC decoding is $\mathcal O (2d_{v}^{\text{ldpc}}L_{c}+(J(2d_{v}^{\text{ldpc}}+1)(L_{c}-L_{b})))$, where $L_{c} $ and $L_{b}$ denote the number of coded bits and information bits, respectively \cite{LinyunSCMA}.  The total complexities of the Turbo like IDD receiver and the proposed SSG-EPA are $N_{{\mathrm {iter}}}^{{\mathrm {out}}}  ({N_{{\mathrm {iter}}}^{{\mathrm {EPA}}}  {\mathcal{ O}}({{\mathrm {EPA}}}) + N_{{\mathrm {iter}}}^{{\mathrm {LDPC}}}  {\mathcal{ O}}({{\mathrm {LDPC}}})})$ and ${N_{{\mathrm {iter}}}}  ({\mathcal {O}({{\mathrm {EPA}}}) + \mathcal {O}({{\mathrm {LDPC}}})})$, respectively.

  \begin{figure}
     \centering
     \includegraphics[width=1\linewidth]{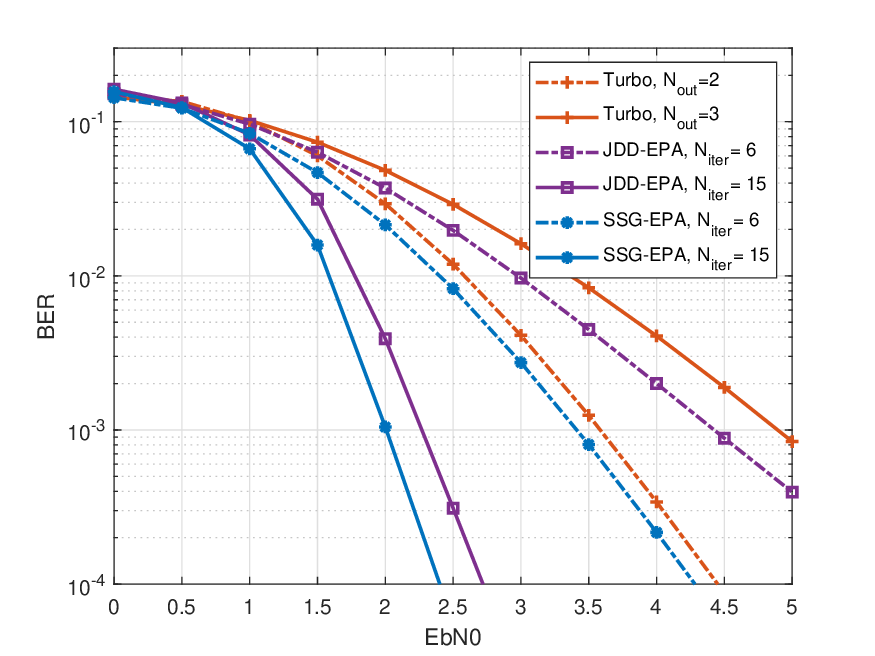}
     \caption{BER comparison of different schemes  for $N_t =1$ and $N_r=2$.}
     \label{nt1}
 \end{figure}

 \begin{figure}
     \centering
     \includegraphics[width=1\linewidth]{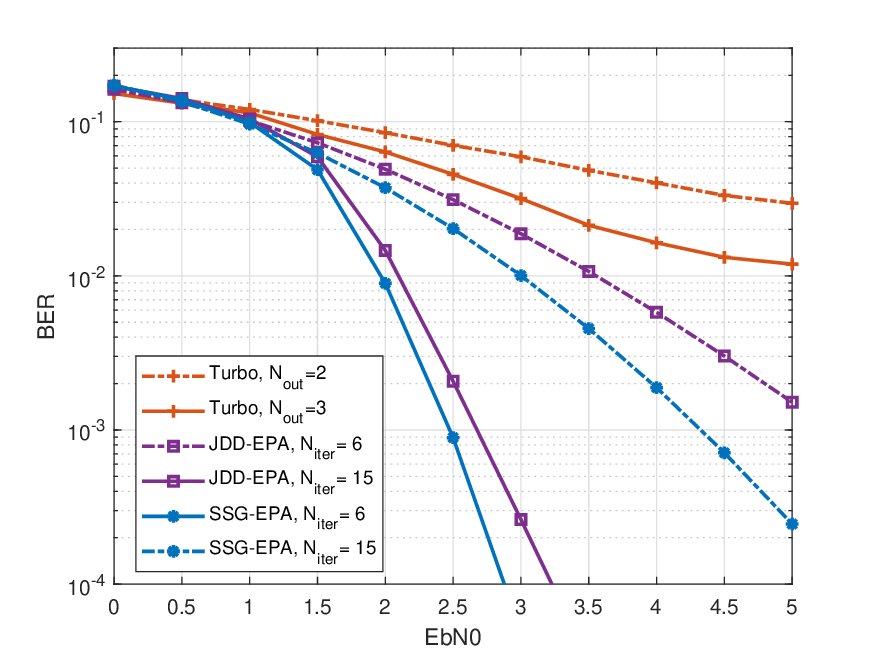}
     \caption{BER comparison of different schemes for $N_t =2$ and $N_r=2$.}
     \label{nt2}
 \end{figure}

\section{Simulation Results}

This section compares the error rate performance of the proposed SSG-EPA with the bench-marking schemes. Specifically, we consider  an uplink SCMA system where the BS equipped $N_r =2$ receive antennas and servers $J=6$ users over $K=4$ orthogonal resources. The codebook size is $M=4$ and the signal space diversity  codebook in \cite{SSDSCMA} is used. The   LDPC code  with the number of information bits and  the coded bits given by $  200$ and $ 400$, respectively, and rate of $1/2$ is used. Therefore, each LDPC frame consists   of $N_{\text{SCMA}}$ SCMA symbols for each user. The maximum number of iterations of the JSG-EPA and JDD-EPA receiver is $N_{\text{iter}}= 15$. For the turbo like IDD-EPA receiver, the outer loop iterations are $N_{\text{iter}}^{\text{out}}= 3$, and the number of  inner loop iterations for both EPA and LDPC is given by $N_{{\mathrm {iter}}}^{{\mathrm {LDPC}}} =N_{{\mathrm {iter}}}^{{\mathrm {EPA}}}  =5$.

Fig. \ref{nt1} shows the coded BER performances of the proposed SSG-EPA, the JDD-EPA and the IDD-EPA receivers with each user equipped a single antenna. As can be seen from the figure, the proposed SSG-EPA achieves the best BER performance among the bench-marking schemes. Specifically, the proposed SSG-EPA achieves about  $0.2$ dB and $1.5$ dB than that of the JDD-EPA and the IDD-EPA receivers for $N_{\text{iter}}= 15$, respectively.  Fig. \ref{nt2} shows the BER comparison of different schemes for $N_t=2$. It should be noted that the same codebook is applied to two transmit antennas. Again, the proposed SSG-EPA still achieves the best BER performance. Both SSG-EPA and JDD-EPA outperform the  IDD-EPA receiver significantly. As the two transmit antennas are employed for multiplexing, the equivalent system is an overloaded system, i.e., $JN_t > KN_r$. In this case, the performance of the EPA with   inner iterations degrades significantly, leading to a performance deteriorate of    the IDD-EPA receiver.   Moreover,   the JSG-EPA and JDD-EPA for $N_t=2$  show about $0.5$ dB performance degradation at $ \text{BER} =10^{-4}$ of  $N_{\text{iter}}=15$ compared to  that of  $N_t=1$ in Fig. \ref{nt1}.    

\begin{figure}
     \centering
     \includegraphics[width=1\linewidth]{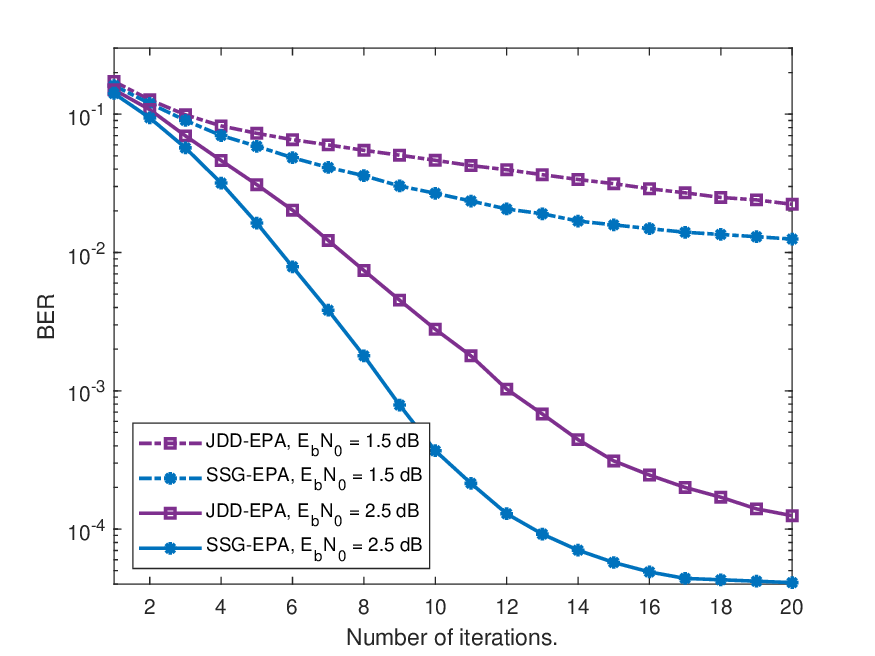}
     \caption{BER performance v.s. number of iterations.}
     \label{Con}
 \end{figure}

Fig. \ref{Con} shows the coded BER performance for different number of iterations of the proposed SSG-EPA and JDD-EPA for $N_t=1$ and $N_r=2$. As can be seen from the figure,   JDD-EPA and SSG-EPA need about $20$   and $16$ iterations for convergence at the $\text{E}_\text{b}/\text{N}_0 = 2.5$ dB, respectively.

\section{Conclusion}

In this paper, we   proposed a SSG-EPA for uplink  MIMO-SCMA. By   reformulating the codebook with a linear encoding model,  the  VNs of SCMA can be transformed from symbol-level   to bit-level VNs. Leveraging this transformation,   w  a SSG has been proposed by emerging  the sparse graphs of SCMA and LDPC into a integrated one.  In addition,  we have  investigated the  use of EPA  in the SSG for low-complexity detection, in resulting in the SSG-EPA receiver.  Notably, the proposed SSG-EPA can update the belief message in an one-shot manner. Namely,   both SCMA   and LDPC  graphs are updated simultaneously, eliminating the use of interleavers, de-interleavers, symbol-to-bit, and bit-to-symbol LLR transformations. 
Simulation results showed that the proposed SSG-EPA achieves better error rate performance than the state-of-the-art schemes.

\ifCLASSOPTIONcaptionsoff
  \newpage
\fi



%
\bibliography{ref} 
\bibliographystyle{IEEEtran}

%








\end{document}